\documentstyle[12pt]{article}

\begin{document}

\author{C. Bizdadea\thanks{%
e-mail address: bizdadea@hotmail.com}, I. Negru and S. O. Saliu\thanks{%
e-mail addresses: osaliu@central.ucv.ro or odile\_saliu@hotmail.com} \\
Department of Physics, University of Craiova\\
13 A. I. Cuza Str., Craiova RO-1100, Romania}
\title{Irreducible Hamiltonian approach to the Freedman-Townsend model}
\maketitle

\begin{abstract}
The irreducible BRST symmetry for the Freedman-Townsend model is derived.
The comparison with the standard reducible approach is also addressed.

\end{abstract}

\section{Introduction}

It is well-known that the Hamiltonian BRST formalism \cite{1}--\cite{5}
stands for one of the strongest and most popular quantization methods for
theories with first-class constraints. In the irreducible context the ghosts
can be interpreted like one-forms dual to the vector fields corresponding to
the first-class constraints. This geometrical interpretation fails within
the reducible framework due to the fact that the vector fields are no longer
independent, hence they cannot form a basis. The redundant behaviour
generates the appearance of ghosts with ghost number greater than one,
traditionally called ghosts for ghosts, of their canonical conjugated
momenta, named antighosts, and, in the meantime, of a pyramidal non-minimal
sector. The ghosts for ghosts ensure a straightforward incorporation of the
reducibility relations within the cohomology of the exterior derivative
along the gauge orbits, while their antighosts are required in order to kill
the higher resolution degree nontrivial co-cycles from the homology of the
Koszul-Tate differential. Among the reducible systems, the Freedman-Townsend
model \cite{17} plays a special role due on the one hand to its link with
Witten's string theory \cite{18}, and, on the other hand, to its equivalence
to the non-linear $\sigma $-model \cite{17}. This model was approached
within the antifield-BRST framework \cite{19}--\cite{23} and only partially
at the Hamiltonian BRST level \cite{19}, but in both settings was studied
along an on-shell reducible context.

The purpose of this paper consists in proving that it is possible to
quantize the Freedman-Townsend model, which is an example of on-shell
first-stage reducible Hamiltonian theory, in the framework of an irreducible
Hamiltonian BRST procedure. The idea of replacing reducible systems by some
irreducible ones is not new. At the Hamiltonian level, this idea appears in 
\cite{5} and \cite{Banerjee}, but it hasn't been either consistently
developed, or applied so far to the quantization of reducible Hamiltonian
systems. Our treatment mainly relies on the replacement of the on-shell
first-stage reducible Hamiltonian model by an irreducible one, and on the
subsequent quantization of the resulting theory in the Hamiltonian BRST
context. The irreducible Hamiltonian system is completely derived from the
requirement that under a suitable redefinition of the antighost number one
antighosts all the antighost number one co-cycles of the reducible
Koszul-Tate differential should identically vanish. This further prevents
the appearance of antighosts with antighost number two. Moreover, the
reducible and irreducible systems possess the same physical observables,
hence the zeroth order cohomological groups of the corresponding BRST
operators coincide. This enables us to replace the BRST symmetry for the
starting reducible theory by that for the irreducible system. As a
consequence of our irreducible approach to the Freedman-Townsend model the
ghosts for ghosts are absent. Thus, the three-ghost coupling term is
discarded from the gauge-fixed action, and the corresponding gauge-fixed
Lagrangian BRST symmetry becomes off-shell nilpotent in our procedure.

The comparison between the reducible analysis and our irreducible treatment
in the case of this model is instructive at emphasizing some interesting
aspects revealed by our formalism. For instance, within the reducible
approach to this model the ghosts for ghosts are bosonic and display
massless scalar field propagators, hence they are fields with correct
spin-statistics relations. On the other hand, it is well-known that in
quantum field theory the ghosts do not describe physical particles, such
that it appears necessary to recover `wrong' spin-statistics relations in
connection with these fields. Our method presents the nice feature that it
restores this required type of `wrong' relations. Another interesting
feature is that the emerging gauge-fixed action is manifestly Lorentz
covariant and displays a simpler form than the one inferred within the
reducible procedure, allowing thus a more straightforward perturbative
approach. In the meantime, our irreducible analysis can be applied to
investigating the possible consistent couplings that can be introduced among
a system of free two-form gauge fields. The last problem was studied in \cite
{23} accordingly the reducible background, being shown that in four
dimensions the Freedman-Townsend interaction vertex defines the only
consistent interaction that deforms nontrivially the gauge transformations
of free two-forms. It is possible that an irreducible approach to this
matter will reveal new aspects or enlighten other features of the already
known results. All these considerations motivate the necessity of an
irreducible analysis for the Freedman-Townsend model.

Our paper is structured in four sections. In Section 2 we realize the
irreducible quantization of the Freedman-Townsend model. Initially, we
derive an irreducible theory associated with the starting reducible model by
means of some homological ideas. Subsequently we prove that we can
substitute the Hamiltonian BRST quantization of the original redundant model
by that of the irreducible theory. In the final part of this section we
deduce the path integral for the irreducible system in the context of the
Hamiltonian BRST quantization, which is found manifestly Lorentz covariant.
In Section 3 we realize the comparison between the gauge-fixed action
derived in Section 2 and the usual gauge-fixed action of the
Freedman-Townsend model obtained in the reducible antifield context. Section
4 ends the paper with some conclusions.

\section{Irreducible treatment of the Freedman-Townsend model}

In this section we construct the path integral for the Freedman-Townsend
model by using an irreducible Hamiltonian BRST procedure. We start with the
canonical analysis and derive some irreducible first-class constraints
associated with the reducible ones following homological arguments. Next, we
obtain the irreducible BRST symmetry associated with the above mentioned
irreducible constraints and reveal its relationship with the standard
reducible BRST symmetry via proving that the physical observables
corresponding to the irreducible and reducible theories coincide. This makes
permissible the replacement of the Hamiltonian BRST quantization for the
original reducible model by that of the irreducible system. Finally, we
deduce the path integral of the irreducible theory, which is found
manifestly Lorentz covariant.

\subsection{Canonical analysis of the model}

We begin with the Lagrangian action of the Freedman-Townsend model \cite{17} 
\begin{equation}
\label{1}S_0^L\left[ A_\mu ^a,B_a^{\mu \nu }\right] =\frac 12\int d^4x\left(
-B_a^{\mu \nu }F_{\mu \nu }^a+A_\mu ^aA_a^\mu \right) ,
\end{equation}
where $B_a^{\mu \nu }$ stands for a set of antisymmetric tensor fields, and
the field strength of $A_\mu ^a$ reads as $F_{\mu \nu }^a=\partial _\mu
A_\nu ^a-\partial _\nu A_\mu ^a-f_{bc}^aA_\mu ^bA_\nu ^c$. The canonical
analysis of this theory outputs the constraints 
\begin{equation}
\label{2}\Phi _i^{(1)a}\equiv \epsilon _{0ijk}\pi ^{jka}\approx 0,\;\Phi
_i^{(2)a}\equiv \frac 12\epsilon _{0ijk}\left( F^{jka}-\left( D^{\left[
j\right. }\right) _{\;\;b}^a\pi ^{\left. k\right] 0b}\right) \approx 0,
\end{equation}
\begin{equation}
\label{3}\chi _i^{(1)a}\equiv \pi _{0i}^a\approx 0,\;\chi _i^{(2)a}\equiv
\pi _a^i+B_a^{0i}\approx 0,\;\chi _a^{(1)}\equiv \pi _a^0\approx 0,
\end{equation}
\begin{equation}
\label{4}\chi _a^{(2)}\equiv A_a^0+f_{ab}^cB_c^{0i}\pi _{0i}^b+\left(
D_i\right) _a^{\;\;b}\pi _b^i\approx 0,
\end{equation}
and the first-class Hamiltonian%
\begin{eqnarray}\label{5}
& &H=\int d^3x\left( \frac 12B_a^{ij}\left( F_{ij}^a-\left( D_{\left[ i\right.
}\right) _{\;\;b}^a\pi _{\left. j\right] 0}^b\right) -\frac 12A_\mu
^aA_a^\mu -\right. \nonumber \\ 
& &\left. A_0^a\left( \left( D_i\right) _a^{\;\;b}\pi
_b^i+f_{ab}^cB_c^{0i}\pi _{0i}^b\right) -A_a^i\left( \pi
_{0i}^a-\partial _i\pi _0^a\right) \right) .
\end{eqnarray}
The symbol $\left[ ij\right] $ appearing in (\ref{5}) signifies the
antisymmetry with respect to the indices between brackets. In the above the
notations $\pi _a^\mu $ and $\pi _{\mu \nu }^a$ denote the momenta
respectively conjugated in the Poisson bracket to the fields $A_\mu ^a$ and $%
B_a^{\mu \nu }$, while the covariant derivatives are defined by $\left(
D_i\right) _{\;\;b}^a=\delta _{\;\;b}^a\partial _i+f_{bc}^aA_i^c$ and $%
\left( D_i\right) _b^{\;\;a}=\delta _b^{\;\;a}\partial _i-f_{bc}^aA_i^c$%
. By computing the Poisson brackets between the constraint functions (\ref{2}%
--\ref{4}) we find that (\ref{2}) are first-class and (\ref{3}--\ref{4})
second-class. In addition, the functions $\Phi _i^{(2)a}$ from (\ref{2}) are
on-shell first-stage reducible 
\begin{equation}
\label{6}\left( \left( D^i\right) _{\;\;b}^a+f_{bc}^a\pi ^{0ic}\right)
\Phi _i^{(2)b}=-\epsilon ^{0ijk}f_{bc}^a\chi _i^{(1)b}\left( D_j\right)
_{\;\;d}^c\chi _k^{(1)d}\approx 0.
\end{equation}
In order to deal with the Hamiltonian BRST formalism, it is useful to
eliminate the second-class constraints with the help of the Dirac bracket 
\cite{Dirac} built with respect to themselves. By passing to the Dirac
bracket, the constraints (\ref{3}--\ref{4}) can be regarded as strong
equalities with the help of which we can express $A_0^a$, $\pi _a^0$, $\pi
_a^i$ and $\pi _{0i}^a$ in terms of the remaining fields and momenta, such
that the independent `co-ordinates' of the reduced phase-space are $A_i^a$, $%
B_a^{0i}$, $B_a^{ij}$ and $\pi _{ij}^a$. The non-vanishing Dirac brackets
among the independent components are expressed by 
\begin{equation}
\label{7}\left[ B_a^{0i}\left( x\right) ,A_j^b\left( y\right) \right]
_{x^0=y^0}^{*}=\delta _a^{\;\;b}\delta _{\;\;j}^i\delta ^3\left( {\bf x}-%
{\bf y}\right) ,
\end{equation}
\begin{equation}
\label{8}\left[ B_a^{ij}\left( x\right) ,\pi _{kl}^b\left( y\right) \right]
_{x^0=y^0}^{*}=\frac 12\delta _a^{\;\;b}\delta _{\;\;k}^{\left[ i\right.
}\delta _{\;\;l}^{\left. j\right] }\delta ^3\left( {\bf x}-{\bf y}\right) .
\end{equation}
In terms of the independent fields, the first-class constraints and
first-class Hamiltonian respectively take the form 
\begin{equation}
\label{9}\gamma _i^{(1)a}\equiv \epsilon _{0ijk}\pi ^{jka}\approx
0,\;G_i^{(2)a}\equiv \frac 12\epsilon _{0ijk}F^{jka}\approx 0,
\end{equation}
\begin{equation}
\label{10}H^{\prime }=\frac 12\int d^3x\left(
B_a^{ij}F_{ij}^a-A_i^aA_a^i+\left( \left( D^i\right)
_{\;\;b}^aB_{0i}^b\right) \left( D_j\right) _a^{\;\;c}B_c^{0j}\right) \equiv
\int d^3x\,h^{\prime },
\end{equation}
while the reducibility relations 
\begin{equation}
\label{11}\left( D^i\right) _{\;\;b}^aG_i^{(2)b}\equiv
Z_{\;\;b}^{ia}G_i^{(2)b}=0,
\end{equation}
hold off-shell in this case. Moreover, the first-class constraints (\ref{9})
remain abelian in terms of the Dirac bracket. In the sequel we work with the
theory based on the reducible first-class constraints (\ref{9}), the
first-class Hamiltonian (\ref{10}) in the context of the Dirac bracket
defined by (\ref{7}--\ref{8}).

\subsection{Irreducible constraints. Irreducible Hamiltonian BRST symmetry}

The Hamiltonian BRST symmetry for our reducible first-class model $%
s_R=\delta _R+\sigma _R+\cdots $ contains two crucial differentials. The
Koszul-Tate differential $\delta _R$ realizes an homological resolution of
smooth functions defined on the surface (\ref{9}), while the model of
longitudinal derivative $\sigma _R$ takes into account the gauge
invariances. The main property of $\delta _R$, the acyclicity, is gained via
introducing some new fields, called antighosts, and denoted by ${\cal P}%
_{(1)i}^a$, ${\cal P}_{(2)i}^a$ and ${\cal P}^a$. The first two sets of
antighosts are fermionic and possess the antighost number one, while the
last set is bosonic and displays the antighost number two. The standard
definitions of $\delta _R$ on the Koszul-Tate generators are given by 
\begin{equation}
\label{12}\delta _Rz^A=0,\;\delta _R{\cal P}_{(1)i}^a=-\gamma _i^{(1)a}, 
\end{equation}
\begin{equation}
\label{13}\delta _R{\cal P}_{(2)i}^a=-G_i^{(2)a}, 
\end{equation}
\begin{equation}
\label{14}\delta _R{\cal P}^a=-\left( D^i\right) _{\;\;b}^a{\cal P}%
_{(2)i}^b, 
\end{equation}
where $z^A$ can be any of the reduced phase-space `co-ordinates'. The
antighosts ${\cal P}^a$ are required by the acyclicity of $\delta _R$ at
antighost number one. Indeed, from (\ref{11}) and (\ref{13}) we find that 
\begin{equation}
\label{15}\mu ^a\equiv \left( D^i\right) _{\;\;b}^a{\cal P}_{(2)i}^b, 
\end{equation}
are nontrivial co-cycles, which are restored trivial with the help of (\ref
{14}).

The basic idea of our irreducible approach consists in redefining the
antighosts ${\cal P}_{(2)i}^a$ in such a way that the new co-cycles of the
type (\ref{15}) vanish identically. If we solve this problem, the antighosts 
${\cal P}^a$ will be discarded from the theory, hence we get an irreducible
Hamiltonian system. In view of this aim, we perform the transformation 
\begin{equation}
\label{16}{\cal P}_{(2)i}^a\rightarrow \tilde{{\cal P}}_{(2)i}^a=N_{\;%
\;bi}^{aj}{\cal P}_{(2)j}^b, 
\end{equation}
where $N_{\;\;bi}^{aj}$ are some functions that may involve the $z^A$'s
taken such that 
\begin{equation}
\label{17}\left( D^i\right)
_{\;\;b}^aN_{\;\;ci}^{bj}=0,\;N_{\;\;bi}^{aj}G_j^{(2)b}=G_i^{(2)a}. 
\end{equation}
Multiplying (\ref{13}) by $N_{\;\;aj}^{bi}$ and taking into account (\ref{16}%
--\ref{17}) we find 
\begin{equation}
\label{18}\delta \tilde{{\cal P}}_{(2)i}^a=-G_i^{(2)a}, 
\end{equation}
which further yield the co-cycles $\nu ^a\equiv \left( D^i\right) _{\;\;b}^a%
\tilde{{\cal P}}_{(2)i}^b$ that vanish identically due to the former
relations in (\ref{17}), hence (\ref{18}) describe an irreducible theory.
The vanishing of these co-cycles induces the removal of the antighost number
two antighosts ${\cal P}^a$ from the antighost spectrum, so it is natural to
replace the notation $\delta _R$ by $\delta $ in (\ref{18}) in order to
emphasize that these relations correspond to an irreducible system. If we
take 
\begin{equation}
\label{19}N_{\;\;bi}^{aj}=\delta _{\;\;b}^a\delta _{\;\;i}^j-\left(
D_i\right) _{\;\;c}^a\bar M_{\;\;d}^c\left( D^j\right) _{\;\;b}^d, 
\end{equation}
with $\bar M_{\;\;d}^c$ the inverse of $M_{\;\;d}^c=\left( D_i\right)
_{\;\;b}^c\left( D^i\right) _{\;\;d}^b$, the relations (\ref{17}) are
automatically fulfilled. It is clear that $M_{\;\;d}^c$ is invertible
because our model is first-stage reducible, i.e., the equations $\left(
D_i\right) _{\;\;b}^au^b\approx 0$ possess only trivial solutions. The
concrete form of $\bar M_{\;\;d}^c$ is not important in the context of our
analysis, but only its existence in principle. Substituting (\ref{19}) in (%
\ref{18}) we arrive at the relations 
\begin{equation}
\label{20}\delta \left( {\cal P}_{(2)i}^a-\left( D_i\right) _{\;\;c}^a\bar
M_{\;\;d}^c\left( D^j\right) _{\;\;b}^d{\cal P}_{(2)j}^b\right)
=-G_i^{(2)a}. 
\end{equation}
The last formulas describe an irreducible theory underlying some irreducible
first-class constraints to be further determined. In this light we add the
supplementary bosonic canonical pairs $\left( \varphi _a,\pi ^a\right) $
equal in number with the number of the reducibility relations (\ref{11}),
with $\pi ^a$ the non-vanishing solutions to the equations 
\begin{equation}
\label{21}M_{\;\;b}^a\pi ^b=\delta \left( \left( D^j\right) _{\;\;b}^a{\cal P%
}_{(2)j}^b\right) . 
\end{equation}
Due to the invertibility of $M_{\;\;b}^a$, the system (\ref{21}) possesses
non-vanishing solutions if and only if $\delta \left( \left( D^j\right)
_{\;\;b}^a{\cal P}_{(2)j}^b\right) \neq 0$. Thus, the non-vanishing
solutions enforce the irreducibility as $\mu ^a$ given by (\ref{15}) are not
co-cycles in this case. Obviously, the opposite situation (i.e., the trivial
solutions $\pi ^b=0$) leads to the equations $\delta \left( \left(
D^j\right) _{\;\;b}^a{\cal P}_{(2)j}^b\right) =0$, that reveal the
appearance of the co-cycles (\ref{15}), and thus of the reducibility. On
behalf of (\ref{21}) the relations (\ref{20}) can be written under the form 
\begin{equation}
\label{22}\delta {\cal P}_{(2)i}^a=-G_i^{(2)a}+\left( D_i\right)
_{\;\;b}^a\pi ^b. 
\end{equation}
Relations (\ref{22}) together with $\delta z^A=0$, $\delta {\cal P}%
_{(1)i}^a=-\gamma _i^{(1)a}$ completely define an irreducible Koszul-Tate
complex based on the irreducible first-class constraints 
\begin{equation}
\label{23}\gamma _i^{(1)a}\equiv \epsilon _{0ijk}\pi ^{jka}\approx
0,\;\gamma _i^{(2)a}\equiv G_i^{(2)a}-\left( D_i\right) _{\;\;b}^a\pi
^b\approx 0. 
\end{equation}
In this manner we associated an irreducible theory based on the irreducible
first-class constraints (\ref{23}) with the initial redundant model. It is
simple to see that the new constraints remain abelian. It is well-known that
the Lagrangian action (\ref{1}) of the original reducible model is invariant
under some Lorentz covariant gauge transformations \cite{19}--\cite{23}. At
the Hamiltonian level, the covariant behaviour of the original gauge
transformations is basically ensured by the presence, besides the gauge
parameters corresponding to the first-class constraints (\ref{2}), of the
supplementary gauge parameters associated with the first-stage reducibility
functions appearing in the left-hand side of (\ref{6}). On the contrary, the
gauge parameters available within the Hamiltonian context of our irreducible
system are fewer than the original ones due to the lack of reducibility.
Thus, we find ourselves in the position to infer some non-covariant
transformations at the level of the Lagrangian action underlying the
irreducible theory. In order to surpass this inconvenient and, moreover, to
derive a manifestly covariant path integral for the irreducible theory it is
necessary to enhance the field and constraint spectrum in such a way to
preserve the irreducibility. A natural and also simple manner of realizing
this scope is to enlarge the phase-space with the canonical bosonic pairs $%
\left( \varphi _a^{(1)},\pi ^{(1)a}\right) $, $\left( \varphi _a^{(2)},\pi
^{(2)a}\right) $, which we set to be constrained by 
\begin{equation}
\label{25}\gamma ^{\prime (1)a}\equiv -\pi ^{(1)a}\approx 0, 
\end{equation}
\begin{equation}
\label{26}\gamma ^{(2)a}\equiv -\pi ^{(2)a}\approx 0. 
\end{equation}
The constraints (\ref{23}) and (\ref{25}--\ref{26}) are first-class and
irreducible. It is well known that one can always add a combination of
first-class constraints to a first-class constraint without afflicting the
theory. In this respect, we remark that (\ref{11}) and (\ref{23}) allow us
to express $\pi ^a$ under the form 
\begin{equation}
\label{27}\pi ^a=-\bar M_{\;\;b}^a\left( D^i\right) _{\;\;c}^b\gamma
_i^{(2)c}, 
\end{equation}
such that we can replace (\ref{25}) (and still maintain the irreducibility)
by 
\begin{equation}
\label{28}\gamma ^{(1)a}\equiv \pi ^a-\pi ^{(1)a}\approx 0. 
\end{equation}
So far, we derived the irreducible first-class constraints 
\begin{equation}
\label{29}\gamma _i^{(1)a}\equiv \varepsilon _{0ijk}\pi ^{jka}\approx
0,\;\gamma _i^{(2)a}\equiv \frac 12\epsilon _{0ijk}F^{jka}-\left( D_i\right)
_{\;\;b}^a\pi ^b\approx 0, 
\end{equation}
\begin{equation}
\label{30}\gamma ^{(1)a}\equiv \pi ^a-\pi ^{(1)a}\approx 0,\;\gamma
^{(2)a}\equiv -\pi ^{(2)a}\approx 0. 
\end{equation}
The first-class Hamiltonian in the Dirac bracket defined by (\ref{7}--\ref{8}%
) with respect to the above constraints can be chosen of the type%
\begin{eqnarray}\label{31}
& &\tilde H=\int d^3x\left( \frac 12B_a^{ij}\left( F_{ij}^a+\varepsilon
_{0ijk}\left( D^k\right) _{\;\;b}^a\pi ^b\right) -\frac 12A_i^aA_a^i+\varphi
_a\pi ^{(2)a}+\right. \nonumber \\  
& &\left. \frac 12\left( \left( D_i\right) _a^{\;\;b}B_b^{0i}-f_{ab}^c\left(
\varphi _c\pi ^b-\varphi _c^{(1)}\pi ^b+\varphi _c^{(2)}\pi ^{(2)b}\right)
\right) ^2-\varphi _a^{(2)}\left( D_i\right) _{\;\;b}^a\left( D^i\right)
_{\;\;c}^b\pi ^c\right) \nonumber \\  
& &\equiv \int d^3x\,\tilde h,
\end{eqnarray}
where we employed the notation%
\begin{eqnarray}\label{32}
& &\left( \left( D_i\right) _a^{\;\;b}B_b^{0i}-f_{ab}^c\left( \varphi _c\pi
^b-\varphi _c^{(1)}\pi ^b+\varphi _c^{(2)}\pi ^{(2)b}\right) \right)
^2\equiv \nonumber \\ 
& &\left( \left( D_i\right) _a^{\;\;b}B_b^{0i}-f_{ab}^c\left( \varphi _c\pi
^b-\varphi _c^{(1)}\pi ^b+\varphi _c^{(2)}\pi ^{(2)b}\right) \right) \times  
\nonumber \\
& &\left( \left( D^j\right) _{\;\;d}^aB_{0j}^d-f_{de}^a\left( \pi
^d\varphi ^e-\pi ^d\varphi ^{(1)e}+\pi ^{(2)d}\varphi ^{(2)e}\right) \right)
.
\end{eqnarray}
The irreducible first-class constraints are abelian in the Dirac bracket,
while the remaining gauge algebra relations read as 
\begin{eqnarray}\label{33}
& &\left[ \gamma _i^{(1)a},\tilde H\right] ^{*}=-\gamma _i^{(2)a},\;\left[
\gamma _i^{(2)a},\tilde H\right] ^{*}=-\left( D_i\right) _{\;\;b}^a\gamma
^{(2)b}+\nonumber \\ 
& &f_{bc}^a\left( \left( D^j\right) _{\;\;d}^bB_{0j}^d-f_{de}^b\left(
\pi ^d\varphi ^e-\pi ^d\varphi ^{(1)e}+\pi ^{(2)d}\varphi ^{(2)e}\right)
\right) \gamma _i^{(2)c},
\end{eqnarray}
\begin{eqnarray}\label{34}
& &\left[ \gamma ^{(1)a},\tilde H\right] ^{*}=\gamma ^{(2)a},\;\left[ \gamma
^{(2)a},\tilde H\right] ^{*}=\left( D^i\right) _{\;\;b}^a\gamma _i^{(2)b}+ 
\nonumber \\
& &f_{bc}^a\left( \left( D^j\right) _{\;\;d}^bB_{0j}^d-f_{de}^b\left(
\pi ^d\varphi ^e-\pi ^d\varphi ^{(1)e}+\pi ^{(2)d}\varphi ^{(2)e}\right)
\right) \gamma ^{(2)c}.
\end{eqnarray}
As it will be further evidenced, the gauge algebra (\ref{33}--\ref{34})
ensures the Lorentz covariance of the irreducible approach.

At this point we are in the position to construct the irreducible BRST
symmetry corresponding to the irreducible theory derived so far. The minimal
antighost spectrum of the irreducible Koszul-Tate differential is organized
as 
\begin{equation}
\label{gho1}{\cal P}_\Gamma \equiv \left( {\cal P}_i^{(1)a},{\cal P}%
_i^{(2)a},{\cal P}^{(1)a},{\cal P}^{(2)a}\right) ,
\end{equation}
where all the variables are fermionic of antighost number one, being
associated with (\ref{29}--\ref{30}). Using the standard definitions 
\begin{equation}
\label{ab1}\delta z^A=0,
\end{equation}
\begin{equation}
\label{ab2}\delta {\cal P}_i^{(\Delta )a}=-\gamma _i^{(\Delta )a},\;\delta 
{\cal P}^{(\Delta )a}=-\gamma ^{(\Delta )a},\;\Delta =1,2,
\end{equation}
the Koszul-Tate operator is nilpotent and acyclic. The generators of the
longitudinal derivative along the gauge orbits are given by 
\begin{equation}
\label{gho}\eta ^\Gamma \equiv \left( \eta _a^{(1)i},\eta _a^{(2)i},\eta
_a^{(1)},\eta _a^{(2)}\right) ,
\end{equation}
and are fermionic with pure ghost number one. The definitions of the
longitudinal derivative along the gauge orbits read as 
\begin{equation}
\label{ab3}\sigma z^A=\sum\limits_{\Delta =1}^2\left( \left[ z^A,\gamma
_i^{(\Delta )a}\right] ^{*}\eta _a^{(\Delta )i}+\left[ z^A,\gamma ^{(\Delta
)a}\right] ^{*}\eta _a^{(\Delta )}\right) ,
\end{equation}
\begin{equation}
\label{ab4}\sigma \eta ^\Gamma =0.
\end{equation}
With these definitions at hand, $\sigma $ is strongly nilpotent. Extending $%
\delta $ to the ghosts (\ref{gho}) and $\sigma $ to the antighosts (\ref
{gho1}) through 
\begin{equation}
\label{ab5}\delta \eta ^\Gamma =0,\;\sigma {\cal P}_\Gamma =0,
\end{equation}
the homological perturbation theory \cite{hom1}--\cite{hom2} guarantees that
the irreducible BRST symmetry $s_I=\delta +\sigma $ exists and is nilpotent.
To conclude with, at this moment we managed to derive an irreducible BRST
symmetry associated with the original reducible one.

\subsection{Relation with the reducible BRST symmetry}

In the sequel we establish the relationship between the irreducible and
reducible BRST symmetries discussed in the above. In this respect we prove
that the physical observables of the irreducible theory coincide with those
of the reducible model. Let $F$ be an observable of the irreducible system.
Consequently, it satisfies the equations 
\begin{equation}
\label{si2}\left[ F,\gamma _i^{(1)a}\right] ^{*}\approx 0,\;\left[ F,\gamma
_i^{(2)a}\right] ^{*}\approx 0,
\end{equation}
\begin{equation}
\label{si3}\left[ F,\gamma ^{(1)a}\right] ^{*}\approx 0,\;\left[ F,\gamma
^{(2)a}\right] ^{*}\approx 0,
\end{equation}
where the weak equality holds on the surface defined by (\ref{29}--\ref{30}%
). Using (\ref{27}) the equations $\left[ F,\gamma _i^{(2)a}\right]
^{*}\approx 0$ lead to 
\begin{equation}
\label{x1}\left[ F,G_i^{(2)a}\right] ^{*}\approx 0,
\end{equation}
on the surface (\ref{29}--\ref{30}). Employing (\ref{27}), by direct
computation we find that 
\begin{equation}
\label{x2}\left[ F,\pi ^a\right] ^{*}\approx 0,
\end{equation}
on the same surface. On behalf of (\ref{x2}) it follows that the equations (%
\ref{si3}) are equivalent with 
\begin{equation}
\label{x3}\left[ F,\pi ^{(1)a}\right] ^{*}\approx 0,\;\left[ F,\pi
^{(2)a}\right] ^{*}\approx 0.
\end{equation}
Thus, any observable of the irreducible theory should verify the former
equations in (\ref{si2}) and (\ref{x1}--\ref{x3}) on the surface (\ref{29}--%
\ref{30}).

Next we show that the first-class constraints (\ref{29}--\ref{30}) are
equivalent with 
\begin{equation}
\label{x4}\gamma _i^{(1)a}\approx 0,\;G_i^{(2)a}\approx 0,\;\pi ^a\approx
0,\;\pi ^{(1)a}\approx 0,\;\pi ^{(2)a}\approx 0. 
\end{equation}
Indeed, the constraints $\gamma _i^{(1)a}\approx 0$, and $\pi ^{(2)a}\approx
0$ pertain to both sets, so it is enough to emphasize that the constraints 
\begin{equation}
\label{x5}G_i^{(2)a}\approx 0,\;\pi ^a\approx 0,\;\pi ^{(1)a}\approx 0, 
\end{equation}
are equivalent to 
\begin{equation}
\label{x6}\gamma _i^{(2)a}\approx 0,\;\gamma ^{(1)a}\approx 0. 
\end{equation}
It is obvious that when (\ref{x5}) hold (\ref{x6}) hold, too. The converse
results as follows. Substituting (\ref{27}) in the concrete expression of $%
\gamma _i^{(2)a}$ we arrive at 
\begin{equation}
\label{si5p}G_i^{(2)a}=\left( \delta _{\;\;d}^a\delta _{\;\;i}^j-\left(
D_i\right) _{\;\;b}^a\bar M_{\;\;c}^b\left( D^j\right) _{\;\;d}^c\right)
\gamma _j^{(2)d}\approx 0. 
\end{equation}
From (\ref{27}) and (\ref{si5p}) we directly get that if (\ref{x6}) hold,
then (\ref{x5}) also hold. This proves the equivalence between the
first-class constraints (\ref{29}--\ref{30}) and (\ref{x4}).

By virtue of the above discussion, we have that any observable of the
irreducible theory, which we found that should verify the former equations
in (\ref{si2}) and (\ref{x1}--\ref{x3}) on the surface (\ref{29}--\ref{30}),
will check these equations also on the surface (\ref{x4}). As a consequence,
the observables of the irreducible theory and those of the theory based on
the constraints (\ref{x4}) coincide. Now, we show that the observables of
the system possessing the constraints (\ref{x4}) and the observables
corresponding to the original reducible theory also coincide. We observe
that the surface (\ref{x4}) can be obtained in a trivial manner from $\gamma
_i^{(1)a}\approx 0$, $G_i^{(2)a}\approx 0$ by adding the canonical pairs $%
\left( \varphi _a,\pi ^a\right) $, $\left( \varphi _a^{(1)},\pi
^{(1)a}\right) $, $\left( \varphi _a^{(2)},\pi ^{(2)a}\right) $ and
requiring that their momenta vanish. Thus, the observables of the original
model are unaffected by the introduction of the new canonical pairs. More
precisely, the difference between an observables $F$ of the theory based on
the constraints (\ref{x4}) and one of the original theory $\bar F$ is of the
type $F-\bar F=\lambda _a\pi ^a+\lambda _a^{(1)}\pi ^{(1)a}+\lambda
_a^{(2)}\pi ^{(2)a}$. As any observables that differ through a combination
of first-class constraints can be identified, it follows that the
observables of the original system and of the theory based on the
constraints (\ref{x4}) coincide. In consequence, we proved that the
observables of the theory with the constraints (\ref{x4}) coincide with
those of the irreducible system, as well as with those of the original
model. This leads to the conclusion that the observables of the irreducible
and original reducible models coincide. In turn, this matter has strong
implications at the BRST level.

As we noticed earlier, there exists a consistent Hamiltonian BRST symmetry
satisfying the general grounds of homological perturbation theory \cite{hom1}%
--\cite{hom2} associated with the irreducible system. Comparing the standard
reducible Hamiltonian BRST symmetry corresponding to the Freedman-Townsend
model $s_R$ with that for the irreducible theory $s_I$ from the point of
view of the basic equations underlying the BRST formalism, we have that 
\begin{equation}
\label{si16}s_R^2=0,\;s_I^2=0,\; 
\end{equation}
\begin{equation}
\label{si17}H^0\left( s_R\right) =H^0\left( s_I\right) =\left\{ {\rm %
physical\;observables}\right\} . 
\end{equation}
The above relations enable us to substitute the Hamiltonian BRST
quantization of the Freedman-Townsend model by that of the irreducible
system.

\subsection{Irreducible path integral}

Based on the last conclusion, we pass to the Hamiltonian BRST quantization
of the irreducible first-class theory, which is described by the first-class
constraints (\ref{29}--\ref{30}) and the first-class Hamiltonian (\ref{31}).
The ghost and antighost spectra are respectively given by (\ref{gho}) and (%
\ref{gho1}). At the same time, we add the non-minimal sector 
\begin{equation}
\label{35}\left( P_{(b)a}^i,b_i^a\right) ,\;\left(
P_{(b^1)a}^i,b_i^{(1)a}\right) ,\;\left( P_{(\bar \eta )a}^i,\bar \eta
_i^a\right) ,\;\left( P_{(\bar \eta ^1)a}^i,\bar \eta _i^{(1)a}\right) ,
\end{equation}
where the first two sets of non-minimal fields are bosonic with ghost number
zero, while the last two sets are fermionic, with the $\bar \eta $'s of
ghost number minus one and the $P_{(\bar \eta )}$'s of ghost number one. The
former variables in every set are taken as fields, while the latter
represent their momenta. Under these considerations, the non-minimal BRST
charge reads as 
\begin{equation}
\label{36}\Omega =\int d^3x\left( \sum\limits_{\Delta =1}^2\left( \eta
_a^{(\Delta )i}\gamma _i^{(\Delta )a}+\eta _a^{(\Delta )}\gamma ^{(\Delta
)a}\right) +P_{(\bar \eta )a}^ib_i^a+P_{(\bar \eta ^1)a}^ib_i^{(1)a}\right) ,
\end{equation}
while the BRST-invariant extension of the first-class Hamiltonian $\tilde H$
is expressed by%
\begin{eqnarray}\label{37}
& &\tilde H_B=\tilde H+\int d^3x\left( \eta _a^{(1)i}{\cal P}_i^{(2)a}-\eta
_a^{(1)}{\cal P}^{(2)a}-\eta _a^{(2)}\left( D^i\right) _{\;\;b}^a{\cal P}%
_i^{(2)b}+\right. \nonumber \\ 
& &\eta _a^{(2)i}\left( D_i\right) _{\;\;b}^a{\cal P}^{(2)b}-f_{ab}^c\left(
\eta _c^{(2)i}{\cal P}_i^{(2)b}+\eta _c^{(2)}{\cal P}^{(2)b}\right) \times  
\nonumber \\
& &\times \left( \left( D^j\right) _{\;\;d}^aB_{0j}^d-f_{de}^a\left( \pi
^d\varphi ^e-\pi ^d\varphi ^{(1)e}+\pi ^{(2)d}\varphi ^{(2)e}\right) \right)
- \nonumber \\
& &\left. \frac 12f_{ab}^cf_{de}^a\left( \eta _c^{(2)i}{\cal P}_i^{(2)b}+\eta
_c^{(2)}{\cal P}^{(2)b}\right) \left( \eta _j^{(2)d}{\cal P}^{(2)je}+\eta
^{(2)d}{\cal P}^{(2)e}\right) \right) . 
\end{eqnarray}
Choosing the gauge-fixing fermion 
\begin{eqnarray}\label{38}
& &K=\int d^3x\left( {\cal P}_i^{(1)a}\left( \varepsilon ^{0ijk}\partial
_jB_{0ka}+\partial ^i\varphi _a^{(1)}\right) -\frac 12\varepsilon _{0ijk}%
{\cal P}^{(1)a}\partial ^iB_a^{jk}+\right. \nonumber \\  
& &\left. P_{(b^1)a}^i\left( {\cal P}_i^{(1)a}-\bar \eta _i^a+%
\stackrel{.}{\bar \eta }_i^{(1)a}\right) +P_{(b)a}^i\left( \bar \eta
_i^{(1)a}+\stackrel{.}{\bar \eta }_i^a\right) \right) ,
\end{eqnarray}
and computing the gauge-fixed action, we find after eliminating some
auxiliary variables the path integral 
\begin{equation}
\label{39}Z_K=\int {\cal D}A_i^a{\cal D}B_a^{\mu \nu }{\cal D}\varphi
_a^{(1)}{\cal D}b_\mu ^a{\cal D}\bar \eta _\mu ^a{\cal D}\eta _a^\mu \exp
\left( i\tilde S_K\right) ,
\end{equation}
where 
\begin{eqnarray}\label{40}
& &\tilde S_K=\int d^4x\left( -\dot A_i^aB_a^{0i}+\frac 12A_i^aA_a^i-\frac
12B_a^{ij}F_{ij}^a-\right. \nonumber \\ 
& &\frac 12\left( \left( D_i\right) _a^{\;\;b}B_b^{0i}+f_{ab}^c\left( \left( 
\stackrel{.}{\bar \eta }_i^b-\partial _i\bar \eta _0^b\right) \eta
_c^i-\left( \partial ^\mu \bar \eta _\mu ^b\right) \eta _c^0\right) \right)
^2-\nonumber \\ 
& &\frac 12\partial _{\left[ i\right. }\bar \eta _{\left. j\right] }^a\left(
D^{\left[ i\right. }\right) _a^{\;\;b}\eta _b^{\left. j\right] }-\left(
\partial ^\mu \bar \eta _\mu ^a\right) \left( \dot \eta _a^0+\left(
D_i\right) _a^{\;\;b}\eta _b^i\right) +\nonumber \\ 
& &b_\mu ^a\left( \frac 12\varepsilon ^{\mu \nu \lambda \rho }\partial _\nu
B_{\lambda \rho a}+\partial ^\mu \varphi _a^{(1)}\right) -\nonumber \\ 
& &\left. \left( \stackrel{.}{\bar \eta }_i^a-\partial _i\bar \eta
_0^a\right) \left( \dot \eta _a^i-\left( D^i\right) _a^{\;\;b}\eta
_b^0\right) \right) .
\end{eqnarray}
In deriving the above expressions we performed the identifications 
\begin{equation}
\label{41}b_\mu ^a=\left( \pi ^{(1)a},b_i^a\right) ,\;\bar \eta _\mu
^a=\left( {\cal P}^{(1)a},-\bar \eta _i^a\right) ,\;\eta _a^\mu =\left( \eta
^{(2)a},\eta _i^{(2)a}\right) .
\end{equation}
We remark that we can introduce an auxiliary field $H_0^a$ in the path
integral by means of the relation%
\begin{eqnarray}\label{42}
& &\exp \left( -\frac i2\int d^4x\left( \left( D_i\right)
_a^{\;\;b}B_b^{0i}+f_{ab}^c\left( \left( \stackrel{.}{\bar \eta }%
_i^b-\partial _i\bar \eta _0^b\right) \eta _c^i-\left( \partial ^\mu \bar
\eta _\mu ^b\right) \eta _c^0\right) \right) ^2\right) =\nonumber \\ 
& &\int DH_0^a\exp \left[ i\int d^4x\left( \frac 12H_0^aH_a^0-\right. \right.  
\nonumber \\
& &\left. \left. H_0^a\left( \left( D_i\right)
_a^{\;\;b}B_b^{0i}+f_{ab}^c\left( \left( \stackrel{.}{\bar \eta }%
_i^b-\partial _i\bar \eta _0^b\right) \eta _c^i-\left( \partial ^\mu \bar
\eta _\mu ^b\right) \eta _c^0\right) \right) \right) \right] .
\end{eqnarray}
With the help of the above Gaussian average and by realizing the
identification 
\begin{equation}
\label{43}A_\mu ^a=\left( H_0^a,A_a^i\right) ,
\end{equation}
the path integral (\ref{39}) takes the manifestly covariant form 
\begin{equation}
\label{44}Z_K=\int {\cal D}A_\mu ^a{\cal D}B_a^{\mu \nu }{\cal D}\varphi
_a^{(1)}{\cal D}b_\mu ^a{\cal D}\bar \eta _\mu ^a{\cal D}\eta _a^\mu \exp
\left( iS_K\right) ,
\end{equation}
where the gauge-fixed action $S_K$ reads as%
\begin{eqnarray}\label{45}
& &S_K=S_0^L\left[ A_\mu ^a,B_a^{\mu \nu }\right] +\int d^4x\left( b_\mu
^a\left( \frac 12\varepsilon ^{\mu \nu \lambda \rho }\partial _\nu
B_{\lambda \rho a}+\partial ^\mu \varphi _a^{(1)}\right) -\right.  
\nonumber \\
& &\left. \frac 12\partial _{\left[ \mu \right. }\bar \eta _{\left.
\nu \right] }^a\left( D^{\left[ \mu \right. }\right) _a^{\;\;b}\eta
_b^{\left. \nu \right] }-\left( \partial ^\mu \bar \eta _\mu ^a\right)
\left( D_\nu \right) _a^{\;\;b}\eta _b^\nu \right) .
\end{eqnarray}
In the above $S_0^L\left[ A_\mu ^a,B_a^{\mu \nu }\right] $ is nothing but
the original action (\ref{1}), and, in addition, we adopted the notations 
\begin{equation}
\label{46}\left( D_0\right) _{\;\;b}^a=\delta _{\;\;b}^a\partial
_0+f_{bc}^aH_0^c,\;\left( D_0\right) _a^{\;\;b}=\delta _a^{\;\;b}\partial
_0-f_{ac}^bH_0^c.
\end{equation}
The BRST symmetries of (\ref{45}) are expressed by 
\begin{equation}
\label{ga}sB_{\mu \nu }^a=\varepsilon _{\mu \nu \lambda \rho }\left(
D^\lambda \right) _{\;\;b}^a\eta ^{\rho b},\;sA_\mu ^a=0,\;s\varphi
^{(1)a}=\left( D_\mu \right) _{\;\;b}^a\eta ^{\mu b},
\end{equation}
\begin{equation}
\label{gi}s\eta _\mu ^a=0,\;s\bar \eta _\mu ^a=b_\mu ^a,\;sb_\mu ^a=0.
\end{equation}
This completes the irreducible Hamiltonian BRST treatment of the
Freedman-Townsend model.

\section{Comparison with the standard reducible approach}

In the following we make the comparison between the results obtained in our
irreducible context and those deriving in the standard reducible BRST
approach. In the reducible approach the gauge-fixed action can be brought to
the form%
\begin{eqnarray}\label{rom49}
& &S_\psi =\int d^4x\left( -\frac 12\partial _{\left[ \mu \right. }\bar
C_{\left. \nu \right] a}\left( D^{\left[ \mu \right. }\right)
_{\;\;b}^aC^{\left. \nu \right] b}-\left( \partial _\mu \bar C_a^\mu \right)
\left( D^\nu \right) _{\;\;b}^aC_\nu ^b-\right. \nonumber \\ 
& &\left( \partial ^\mu \bar C_a\right)
\left( D_\mu \right) _{\;\;b}^a{\cal C}%
^b+\frac 18\epsilon ^{\mu \nu \lambda \rho }f_{\;\;bc}^a\partial _{\left[
\mu \right. }\bar C_{\left. \nu \right] a}\partial _{\left[ \lambda \right.
}\bar C_{\left. \rho \right] }^c{\cal C}^b+\nonumber \\ 
& &\left. \left( \frac 12\epsilon ^{\mu \nu \lambda \rho }\partial
_\nu B_{\lambda \rho a}+\partial ^\mu \bar C_a^{\prime }\right) {\cal B}_\mu
^a\right) +S_0^L,
\end{eqnarray}
where $C_\nu ^a$ represent the ghost number one ghosts, and ${\cal C}^a$
signify the ghosts for ghosts. The gauge-fixed BRST symmetries of (\ref
{rom49}) read as 
\begin{equation}
\label{rom49a}s_\psi B_{\mu \nu }^a=\varepsilon _{\mu \nu \lambda \rho
}\left( D^\lambda \right) _{\;\;b}^aC^{\rho b}+f_{\;\;bc}^a\partial _{\left[
\mu \right. }\bar C_{\left. \nu \right] }^c{\cal C}^b,\;s_\psi C_\mu
^a=\left( D_\mu \right) _{\;\;b}^a{\cal C}^b,s_\psi {\cal C}^a=0,
\end{equation}
\begin{equation}
\label{rom49b}s_\psi A_\mu ^a=0,\;s_\psi \bar C_\mu ^a={\cal B}_\mu
^a,\;s_\psi \bar C^a=\partial ^\mu \bar C_\mu ^a,\;s_\psi \bar C^{\prime
a}=\partial ^\mu C_\mu ^a,\;s_\psi {\cal B}_\mu ^a=0.
\end{equation}
The above symmetries are on-shell nilpotent with respect to some of the
fields 
\begin{equation}
\label{rom49c}s_\psi ^2B_{\mu \nu }^a=-\varepsilon _{\mu \nu \lambda \rho
}f_{\;\;bc}^a\frac{\delta S_\psi }{\delta B_{\lambda \rho c}}{\cal C}%
^b,\;s_\psi ^2\bar C^a=-\frac{\delta S_\psi }{\delta \bar C_a^{\prime }}%
,\;s_\psi ^2\bar C^{\prime a}=\frac{\delta S_\psi }{\delta \bar C_a},
\end{equation}
and off-shell nilpotent otherwise. It is obvious that the ghosts for ghosts $%
\left( \bar C_a,{\cal C}^a\right) $ in the gauge-fixed action (\ref{rom49})
possess massless scalar field propagators, hence they output correct
spin-statistics relations, in disagreement with the requirement of not
describing physical particles. In the meantime, we cannot surpass this
inconvenient by integrating over the ghosts for ghosts. This is because the
integration over $\left( \bar C_a,{\cal C}^a\right) $ is cumbersome as the
matrix of their quadratic parts depends on the fields and the structure
constants, such that one cannot eliminate this term from the path integral.
By contrast, the ghosts for ghosts are absent within the gauge-fixed action (%
\ref{45}), so the above disagreement is surpassed within the irreducible
analysis. By performing the identifications 
\begin{equation}
\label{rom50}C_\mu ^a\leftrightarrow \eta _\mu ^a,\;\bar C_\mu
^a\leftrightarrow \bar \eta _\mu ^a,\;\bar C^{\prime a}\leftrightarrow
\varphi ^{(1)a},\;{\cal B}_\mu ^a\leftrightarrow b_\mu ^a,
\end{equation}
among the variables involved with the gauge-fixed actions derived within the
irreducible and reducible approaches, (\ref{45}), respectively, (\ref{rom49}%
), the difference between the two gauge-fixed actions becomes 
\begin{equation}
\label{rom51}S_\psi -S_K=\int d^4x\left( -\left( \partial ^\mu \bar
C_a\right) \left( D_\mu \right) _{\;\;b}^a{\cal C}^b+\frac 18\epsilon ^{\mu
\nu \lambda \rho }f_{\;\;bc}^a\partial _{\left[ \mu \right. }\bar C_{\left.
\nu \right] a}\partial _{\left[ \lambda \right. }\bar C_{\left. \rho \right]
}^c{\cal C}^b\right) .
\end{equation}
We remark that $S_\psi -S_K$ is proportional with the ghosts for ghosts, $%
{\cal C}^b$, which are some essential compounds of the reducible BRST
quantization. Although identified at the level of the gauge-fixed actions,
the fields $\bar C^{\prime a}$ and $\varphi ^{(1)a}$ play different roles
within the two formalisms. More precisely, the presence of $\varphi ^{(1)a}$
within the irreducible model prevents the appearance of the reducibility,
while the $\bar C^{\prime a}$'s represent an effect of the reducibility. In
fact, the Lagrangian effect of introducing the fields $\varphi ^{(1)a}$
resides in forbidding the existence of the zero modes which are present
within the original reducible theory. In consequence, all the ingredients
connected with the zero modes, like the ghosts for ghosts or the non-minimal
pyramid, are no longer involved. In this light, we suggestively call the
fields $\varphi ^{(1)a}$ `antimodes'. The antimodes also contribute to the
difference between the corresponding gauge-fixed BRST symmetries. Indeed,
the absence of the ghosts for ghosts in our irreducible context makes the
gauge-fixed BRST symmetries (\ref{ga}--\ref{gi}) off-shell nilpotent, by
contrast with the reducible situation (see (\ref{rom49c})), and, moreover,
removes the three-ghost coupling term from our irreducible procedure.
Finally, we mention that the number of antimodes is equal with the number of
the pairs $\left( \bar C_a,{\cal C}^a\right) $ generated by the zero modes.
Obviously, neither the ghost for ghost pairs nor the antimodes describe
physical particles and are however governed by correct spin-statistics
relations, but, while the ghosts for ghosts cannot be eliminated from the
path integral by direct integration, the antimodes do not produce this
difficulty due to the possibility of a safely integration over them by
making a Gaussian average in (\ref{45}).

\section{Conclusion}

In this paper we show that the Freedman-Townsend model, which is a typical
example of on-shell first-stage reducible Hamiltonian system, can be
consistently approached along the irreducible Hamiltonian BRST line by
replacing the original reducible Hamiltonian theory with an irreducible one.
The construction of the irreducible system is based on ``killing'' the
initial redundancy at the level of the Koszul-Tate complex. As the physical
observables associated with the irreducible and reducible versions coincide,
the main equations underlying the Hamiltonian BRST formalism make legitimate
the substitution of the reducible BRST symmetry by the irreducible one. The
gauge-fixed action of the irreducible model is derived within the
Hamiltonian BRST context by using an appropriate gauge-fixing fermion.
Finally, we emphasize the comparison between our irreducible procedure and
the standard reducible BRST treatment of the Freedman-Townsend model.

\end{document}